\documentclass[a4paper,11pt]{article}
\usepackage{pos}
\usepackage[backend=bibtex,url=false,doi=false,hyperref,style=numeric,sorting=none,bibencoding=utf8,maxnames=2,uniquename=init,giveninits=true]{biblatex}
\usepackage{braket}
\usepackage{physics}
\usepackage{booktabs}
\usepackage{bbm} 
\usepackage{bm} 
\title{Variance reduction via deflation with local coherence}

\author*[a]{Roman Gruber}
\author[a]{Tim Harris}
\author[a]{Marina Krstić Marinković}

\affiliation[a]{Institut f\"ur Theoretische Physik, ETH Z\"urich, \\ Wolfgang-Pauli-Str. 27, 8093 Z\"urich, Switzerland}

\emailAdd{rgruber@ethz.ch}
\emailAdd{harrist@phys.ethz.ch}
\emailAdd{marinama@ethz.ch}

\abstract{
    In large enough volumes, translation-averaging for quark-line connected
    diagrams reduces the variance inversely proportional to the volume.
    Stochastic estimators which implement translation averaging however
    introduce new sources of fluctuations, which in some cases can be
    relatively large.
    In this work,
   {we explore whether inexact deflation subspaces can be used to improve the precision of the isovector vector correlators}. 
    We perform numerical experiments with $N_\mathrm{f}=2$ non-perturbatively
    $\mathrm O(a)$-improved Wilson fermions and measure the relative
    contribution from the deflation subspace to the central value and the
    corresponding variance.
}

\FullConference{The 40th International Symposium on Lattice Field Theory (Lattice 2023)\\
July 31st - August 4th, 2023\\
Fermi National Accelerator Laboratory\\}

\begin{document}
\maketitle

\section{Introduction}

\newcommand{\Nf}{N_\mathrm{f}}
\newcommand{\Nsrc}{N_\mathrm{src}}
\newcommand{\Nb}{N_\mathrm{b}}
\newcommand{\Ns}{N_\mathrm{s}}

When translation averaging is used, the variance of observables with a small
footprint is expected to decrease inversely proportional to the volume in a
Monte Carlo simulation due to the mass gap of QCD~\cite{Luscher:2017cjh}.
While this idea certainly works out in the master-field
regime~\cite{Giusti:2018cmp}, it has commonly been put into practice for
moderate volumes, too.
For quark-line connected correlation functions, such as the isovector vector
correlator
\begin{align}
    \langle G(x,y_0)\rangle_\mathrm{g}
    &= \langle\tr\{W(x,y_0,x)\}\rangle_\mathrm{g},\qquad
        W(x,y_0,z)=-\tfrac{1}{3}\sum_ka^3\sum_{\bm y}\{S(x,y)\gamma_kS(y,z)\gamma_k\},
    \label{eq:trace}
\end{align}
where $S(x,y)$ denotes the light-quark propagator and
$\langle\cdot\rangle_\mathrm{g}$ 
denotes the average with respect to the effective action after integrating out the fermion fields, the variance shows a suppression proportional to the volume.
This holds true even when integrating both vertices over the spatial volume $L^3$ in a box of linear size $L$.
This can be understood from the expression for the variance for its
translation-averaged counterpart
$G_\mathrm{vol}(x_0,y_0)=\frac{a^3}{L^3}\sum_{\bm{x}}G(x,y_0)$ which is
\begin{align}
    \sigma^2_\mathrm{vol}(x_0-y_0) &=
    \frac{a^3}{L^3}
        \sum_{x'} \delta_{x_0x_0'}
        \Big[
            \langle G(x,y_0)G(x',y_0)\rangle_\mathrm{g}-\langle G(x,y_0)\rangle_\mathrm{g}\langle G(x',y_0)\rangle_\mathrm{g}
        \Big]
    \label{eq:vol}
\end{align}
and noting the sum converges in infinite volume thanks to the extra suppression
due to the quark propagators.
Note also that the variance has no power-law divergences for $x_0\neq y_0$.

Implementing translation averaging is not completely straightforward without
incurring a cost which also increases with the volume, for example if
propagators would be explicitly computed from every lattice site.
One possibility is to use a stochastic estimator for the translation average,
such as the one-end trick~\cite{Michael:1997rk,deDivitiis:1996qx}
\begin{align}
    \label{eq:stochest}
    \mathcal G_\mathrm{vol}(x_0,y_0) &=
    \frac{1}{\Nsrc}\sum_{i=1}^{\Nsrc} \sum_{\alpha=1}^4 \frac{a^3}{L^3}\sum_{\bm x,z}
    \eta_i^{(\alpha) \dagger}(x) W(x,y_0,z)\eta_i^{(\alpha)}(z) \delta_{x_0z_0}
\end{align}
where the dimensionless fields $\eta^{(\alpha)}_{i\beta
c}(x)=\delta^{\alpha}_{\beta}\xi_{ic}(x)$ are defined in terms of i.i.d. random fields
with zero mean and variance
\begin{align}
    \langle \xi_{ic}(x) \xi_{jd}^*(y) \rangle_\xi &=
    \delta_{ij}\delta_{cd}\delta_{xy},
    \label{eq:noise}
\end{align}
and we have chosen the spin-diagonal variant~\cite{ETM:2008zte} which allows
the unbiased estimation of each spin matrix element of $W$ with just $4\Nsrc$
inversions.

The introduction of new auxiliary fields leads to fluctuations, which, in the
case of Gaussian-distributed fields, introduces an extra contribution to the
variance~\cite{Luscher:2010ae}
\begin{align}
    \sigma^2_\mathrm{stoch} = \sigma^2_{\mathcal G} - \sigma^2_\mathrm{vol} &=
    \frac{1}{\Nsrc}\sum_{\alpha,\beta=1}^4\sum_{a,b=1}^3\sum_{\bm x,z}\delta_{x_0z_0}\langle
        W_{\alpha a,\alpha b}(x,y_0,z)W_{b\beta,a\beta}(z,y_0,x)
    \rangle_\mathrm{g}.
    \label{eq:stochvar}
\end{align}
Although its volume-scaling is the same as the exact average in
Eq.~\eqref{eq:vol}, it has different matrix elements which can be relatively
large compared to $\sigma^2_\mathrm{vol}$, potentially spoiling the
variance-reduction from translation averaging.
This is illustrated in Fig.~\ref{fig:final} where we depict estimators for the
variances $\sigma^2_\mathrm{vol}$ (red) and $\sigma^2_\mathrm{stoch} \approx \sigma^2_{\mathcal G}$ (blue) for the
isovector vector correlator showing the large hierarchy between them even when
$\Nsrc$ is large.
In the following we investigate improved estimators based on deflation
techniques.

\section{Variance reduction via deflation}
Natural decompositions of traces such as Eq.~\eqref{eq:trace} can be
constructed starting from a decomposition of the quark propagator
defined in terms of a set of $N$ linear independent fields $\phi_1(x),\ldots,\phi_N(x)$
\begin{align}
    S(x,y) &= \sum_{k,l=1}^N\phi_k(x)(A^{-1})_{kl}\phi_l^\dagger(y) + P_\mathrm{R}S(x,y),
    \label{eq:lohi}
\end{align}
where the first term represents the propagator within the subspace of fields,
and the matrix elements of the so-called little Dirac operator $A$ are defined
explicitly in terms of the Dirac operator $D$
\begin{align}
    A_{kl} &= (\phi_k, D\phi_l).
    \label{eq:lo}
\end{align}
Such operators arise when considering the Schur complement of the propagator in
the subspace.
The remainder may be written in terms of the projected propagator, namely
\begin{align}
    P_\mathrm{R}S(x,y) &= S(x,y) - \sum_{k,l=1}^N\phi_k(x)(A^{-1})_{kl}\phi_l^\dagger(y).
    \label{eq:hi}
\end{align}
Inserting the representation Eq.~\eqref{eq:lohi} into Eq.~\eqref{eq:trace}, the
trace may be decomposed into terms involving only the little propagator
$A^{-1}$ and a remainder.
If the fields are chosen to be exact (left) singular vectors of the Dirac operator,
the former contribution can be computed exactly with translation averaging.
When the remainder is computed stochastically, the extra contribution to the
variance has the same form as Eq.~\eqref{eq:stochvar} with at least two
insertions of $P_\mathrm{R}$.
The efficiency of the deflation for our purposes is thus measured by the
reduction in the variance.
In practice, we note that the dimension of the little propagator is too large
to compute and store directly, and thus its contributions may also need to be
estimated stochastically.
Nevertheless, here we will assume it is cheap enough that the extra
contribution to the variance can be suppressed by a sufficiently large number
of samples $\Nsrc$.

Such low-mode averaging and its variants are common variance-reduction
strategies in lattice QCD and have been known for a long
time~\cite{degrand2004,Giusti_2004}.
However, given that the spectral density increases with the volume, the cost
of achieving a constant efficiency with the volume scales with its square.
Volume scaling e.g. achieved in inexact deflation~\cite{Luescher2007} and multigrid solvers~\cite{brannick2008} profits 
from the local coherence of the low modes of the Dirac operator. 
That is, the low modes projected to local domains can be represented by a small
basis of fields to a high degree of precision.
In this way, only a small number of exact modes are required to span a much
larger deflation subspace and the volume-scaling problem is circumvented, as
outlined in the following section.

\subsection{Domain-decomposed subspaces}
A domain-decomposed subspace of dimension $N$ may be constructed from a much
smaller set of $\Ns$ input fields $\psi_1(x),\ldots,\psi_{\Ns}(x)$ in the
following way.
By decomposing the lattice into a set of $\Nb$ non-overlapping blocks labelled by
$\Lambda$, the input fields are projected to the blocks
\begin{align}
    \psi_k^\Lambda(x) &=
    \begin{cases}
        \psi_k(x) & \textrm{if $x\in\Lambda$},\\
        0 & \textrm{otherwise}
    \end{cases}
    \label{eq:blocks}
\end{align}
and re-orthogonalized afterwards.
After relabelling the $N=\Nb\Ns$ fields one implements the deflation exactly as
described in the previous subsection.
In contrast to the input fields used in the preconditioning of the Dirac
equation~\cite{Luescher2007} which may be fairly rough approximations of the low modes, initialized
for example through a few inverse iterations, we have in mind to
take the input fields as exact (left) singular vectors of the lattice Dirac
operator.
Such domain decompositions have been utilized as a compression
algorithm~\cite{Clark_2018} as well as variance reduction similar to what we
investigate here~\cite{Alcalde_2017}.

\section{Numerical experiments}
\label{sec:numeric}

\begin{table}[t]
\centering
\begin{tabular}{ccccc}
\toprule
Name & $T \times L^3$ & Pion mass & $a$ [fm] & $\#$ configs \\
\midrule
D5d & $48 \times 24^3$ & $439$ MeV & $0.0658(10)$ & 100 \\
F7 & $96 \times 48^3$ & $268$ MeV & $0.0658(10)$ & 100 \\
\bottomrule
\end{tabular}
\caption{Ensembles of $\Nf=2$ non-perturbatively $\mathrm O(a)$-improved Wilson
fermions used in this work.}
\label{tab:ensembles}
\end{table}

In this section, we present some numerical investigations of domain-decomposed
deflation subspaces using $\Nf=2$ non-perturbatively $\mathrm O(a)$-improved
Wilson fermions generated by the Coordinated Lattice Simulations (CLS)
effort~\cite{cls}.
We study two ensembles of representative gauge field configurations
separated widely in Monte Carlo time and whose parameters are listed in
Tab~\ref{tab:ensembles}.
A small number of input fields $\Ns\sim 10-100$ were created on each
configuration using the PRIMME library~\cite{PRIMME} to a relative precision\footnote{$\norm{ D \gamma5 \chi_i - \lambda_i \chi_i }/\norm{\chi_i} < 10^{-12}$.} of
$10^{-12}$.
The stochastic sources $\eta$ were spin-diagonal with $\mathbb{Z}_2 \times \mathbb{Z}_2$ noise.
We report results only for the case of the isovector vector current correlator
as defined in Eq.~\eqref{eq:trace}.

\subsection{Efficiency of domain-decomposed subspaces}
\begin{figure}
\centering
\includegraphics[width=.5\linewidth]{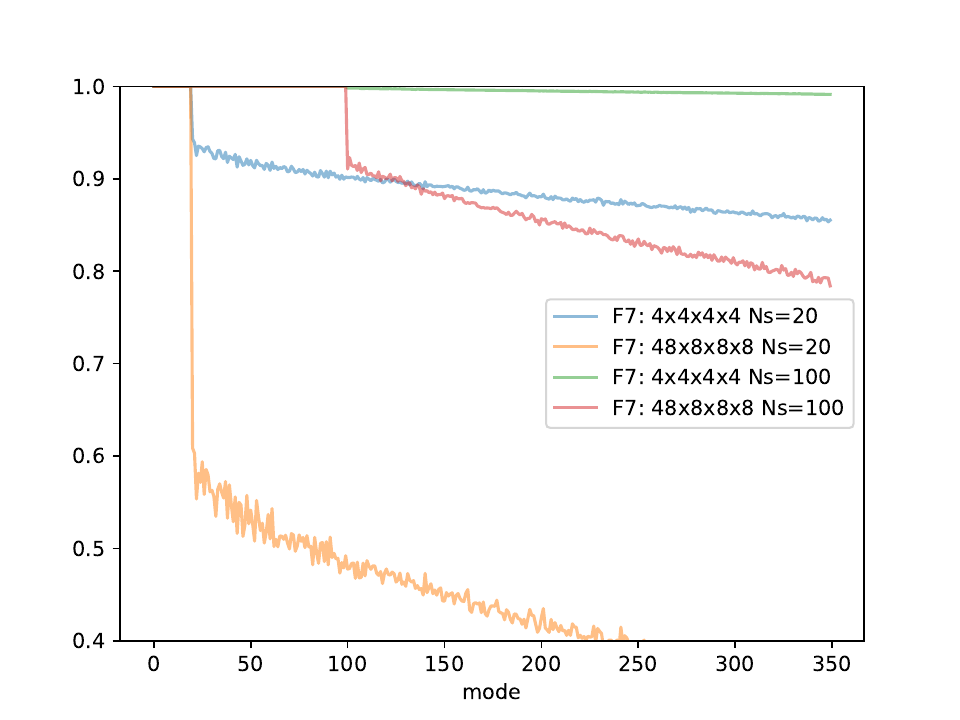}
\caption{Local coherence of low modes. The deviation of the deficits
Eq.~\eqref{eq:def} from unity $1-\epsilon_i$ versus the mode number. The first $\Ns$ modes in
each case are exactly contained in the enlarged subspace by construction and
have zero deficit. The deficits are reduced by decreasing the domain size or by
increasing $\Ns$ the number of input modes.}
    \label{fig:dd_lc}
\end{figure}

In order to illustrate one sense in which the enlarged space of fields
represents the true low-mode subspace well, one can examine the deficits
$\epsilon_i$ of the exact low modes $\chi_i$ which are not contained in the
domain-decomposed subspace
\begin{align}
    \epsilon_i = \norm{\chi_i-\textstyle\sum_{k=1}^N\phi_k(\phi_k,\chi_i)}/\norm{\chi_i}
    \label{eq:def}
\end{align}
In Fig.~\ref{fig:dd_lc}, we display the deficits of the first few hundred exact
low modes on the F7 ensemble for various subspaces created from two choices of
number of input fields $\Ns=20,100$ and two configurations of the domain sizes
namely $4^4$ and $48\times8^3$ in lattice units, which correspond to domains in
the spatial sizes of around $0.25$ and $0.5\,\mathrm{fm}$ respectively.
The subspace size is thus increased by a factor $41472$ or $432$ in each case
by the domain decomposition.
As expected, the deficits are reduced by decreasing the domain size and by
increasing the number of input modes.
However, as explained previously, the precise measure of the efficiency for our
purpose is whether the stochastic variance of the remainder terms is reduced
compared to the undeflated version, which we investigate in the
following subsection.

\subsection{Efficiency of variance reduction}
In this section we test the efficiency of the deflation in reducing the
variance on the individual contributions to the trace in Eq.~\eqref{eq:trace}
for various domain decompositions.
Using the deflation subspace as defined above, we make the decomposition of the
stochastic estimator of Eq.~\eqref{eq:stochest}
\begin{align}
    \mathcal G_\mathrm{vol} &= \mathcal G^\mathrm{lttl}_\mathrm{vol}
    + \mathcal G^\mathrm{rem}_\mathrm{vol}
    \label{eq:decomp}
\end{align}
where the two estimators are defined as follows.
The first contains only the little propagators
\begin{align}
    \nonumber
    \mathcal G^\mathrm{lttl}_\mathrm{vol}(x_0,y_0) &=
    \frac{1}{\Nsrc}\sum_{i=1}^{\Nsrc} \sum_{\alpha=1}^4 \frac{a^6}{L^3}\sum_{\bm x,\bm y,z}\delta_{x_0z_0}\Big[\\
    &\qquad\sum_{klmn=1}^N \eta_i^{(\alpha) \dagger}(x) \phi_k(x) (A^{-1})_{kl} \phi_l^\dagger(y)\gamma_k
    \phi_m(y) (A^{-1})_{mn} \phi^\dagger_n(z)\gamma_k\eta_i^{(\alpha)}(z) \Big]
    \label{eq:lovol}
\end{align}
while the remainder contains a mixed term and a term arising only from the
projected propagator
\begin{align}
    \nonumber
    \mathcal G^\mathrm{rem}_\mathrm{vol}(x_0,y_0) &=
    \frac{1}{\Nsrc} \sum_{i=1}^{\Nsrc} \sum_{\alpha=1}^4 \frac{a^6}{L^3}\sum_{\bm x,\bm y,z}\delta_{x_0z_0}\Big[
        2\sum_{kl=1}^N \eta_i^{(\alpha) \dagger}(x) \phi_k(x) (A^{-1})_{kl} \phi_l^\dagger(y)\gamma_k
            P_\mathrm{R}S(y,z)\gamma_k\eta_i^{(\alpha)}(z) \\
        &\qquad\qquad\qquad\qquad+ \eta_i^{(\alpha) \dagger}(x) P_\mathrm{R}S(x,y)\gamma_k
    P_\mathrm{R}S(y,z)\gamma_k\eta_i^{(\alpha)}(z) \Big].
    \label{eq:hivol}
\end{align}
As explained earlier, we assume that in practice the little propagator is cheap
enough so that $\Nsrc$ can be chosen very large for that term.
We note, however, that the one-end trick estimator is recovered if $\Nsrc$ is
chosen to be the same for both terms with the same noise fields.
It is therefore interesting to examine each term in turn with a fixed number of
$\Nsrc$, keeping in mind that the variance can be effectively eliminated on the
little term at no cost.

\begin{figure}
\centering
\begin{minipage}{.5\textwidth}
  \centering
  \includegraphics[width=\textwidth]{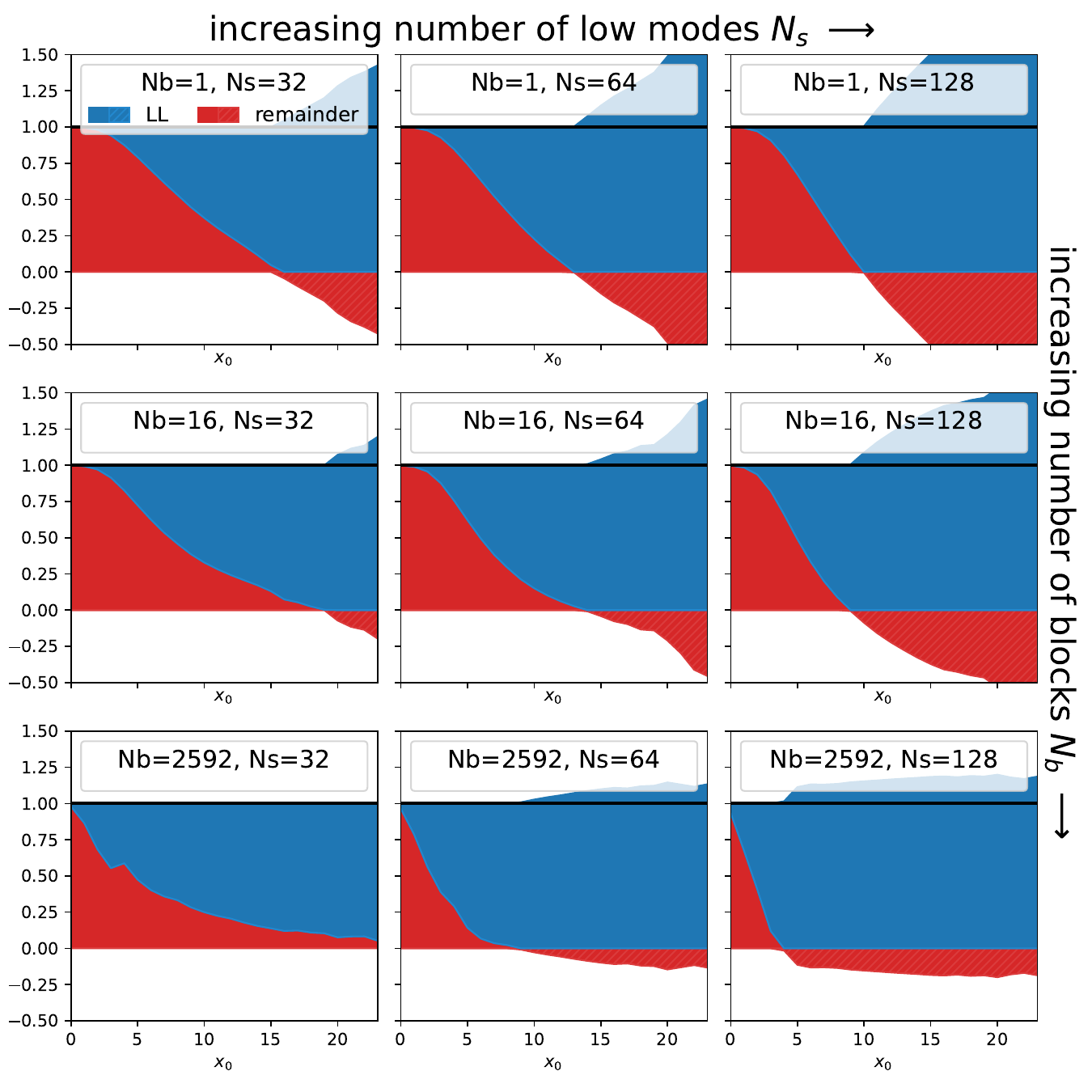}
\end{minipage}%
\begin{minipage}{.5\textwidth}
  \centering
  \includegraphics[width=\textwidth]{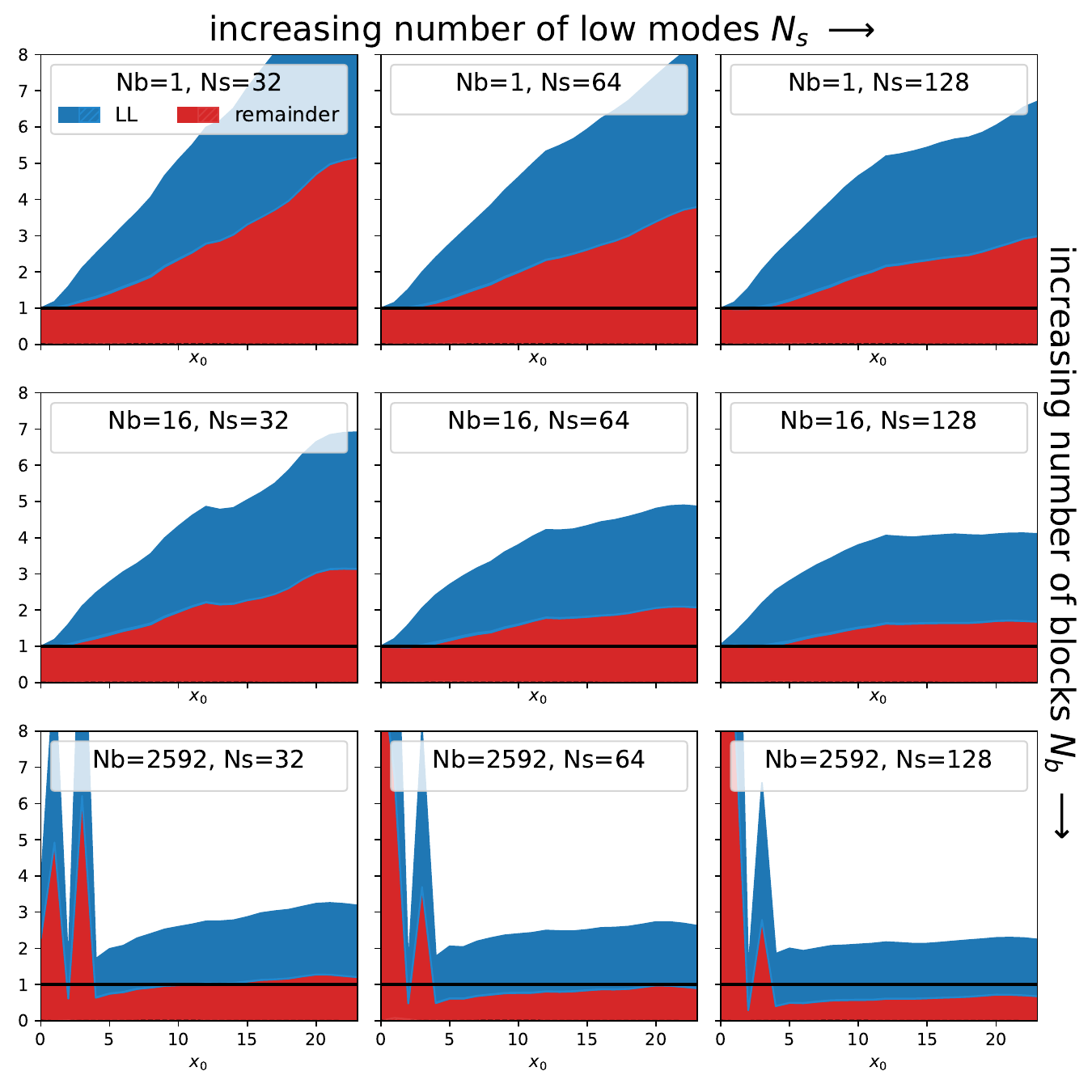}
\end{minipage}

\caption{Left: Relative contributions to the value of the correlator Eq.
    \eqref{eq:trace}, on D5d lattice using different number of low modes $\Ns$
    and different number of blocks $\Nb$ increasing left to right and top to
    bottom respectively. Right: The variance of the little contribution (blue)
    and remainder (red) normalized by the variance of the sum.
    The block sizes ($T \times L^3$) are; 1st row: no blocking
    (corresponding to traditional LMA), 2nd row: $24 \times 12 \times 12 \times 12$,
    3rd row: $4 \times 4 \times 4 \times 4$.}
\label{fig:D5d}
\end{figure}

\begin{figure}
\centering
\begin{minipage}{.5\textwidth}
  \centering
  \includegraphics[width=\textwidth]{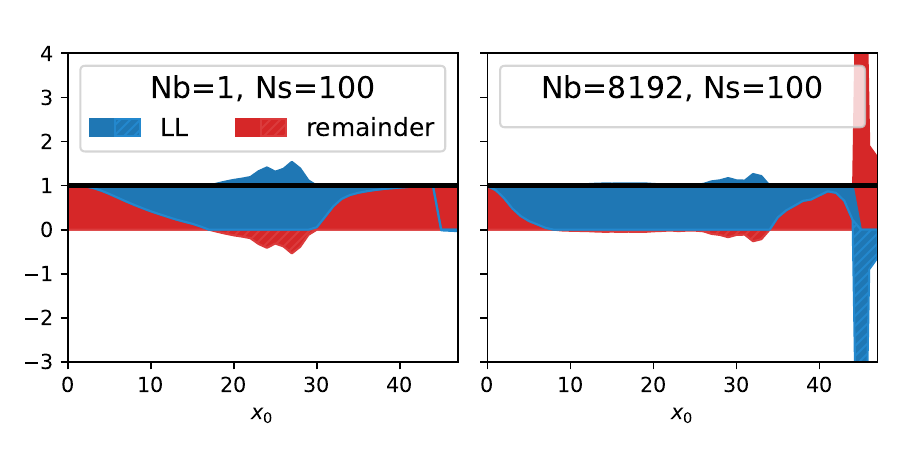}
\end{minipage}%
\begin{minipage}{.5\textwidth}
  \centering
  \includegraphics[width=\textwidth]{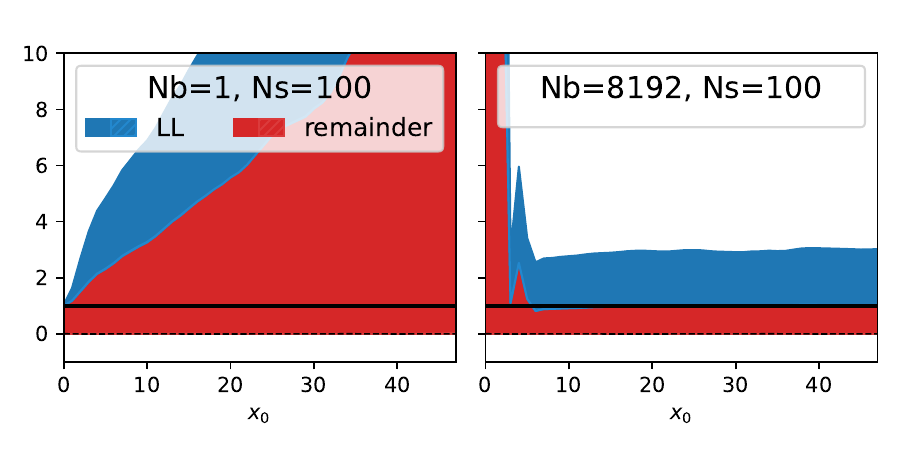}
\end{minipage}

\caption{Same as Fig.~\ref{fig:D5d} for the F7 ensemble. The block sizes ($T \times L^3$) are; left: no blocking (LMA) right: $6 \times 6 \times 6 \times 6$.}
\label{fig:F7}
\end{figure}

In Figs.~\ref{fig:D5d} and~\ref{fig:F7} we show the relative contribution of the
two terms to both the central value and the variance as a function of
$x_0-y_0$.
In the left panels, the relative contribution to the central value of the
little term (blue) and the remainder term (red) is shown normalized by the
sum.
Each panel represents a different choice of parameters, with the number of
blocks $\Nb$ increasing from left to right, and the number of input fields
$\Ns$ increasing from top to bottom.
Note that the remainder term, representing the contribution from the high modes
dominates at small separations but even becomes negative at large separations.

In the right panel, we show the variances of the little term (blue) and the
remainder (red), normalized by the variance of the sum.
The striking feature is the large size of the variance from the remainder term
even at the largest separations.
Except for the largest deflation subspace sizes (bottom right), this
illustrates that the variance on the one-end trick estimator is smaller than
the individual contributions, which demonstrates the high degree of covariance
between the two pieces.
That is, the stochastic variance on the remainder is not guaranteed to be
smaller than the original estimator.
Even if the variance due to the little propagator is eliminated by increasing
the number of samples, the variance on the remainder (red) is larger than
the original estimator (black line) except for the smallest domain size of
$4^4$ for the D5d ensemble (Fig.~\ref{fig:D5d} right, bottom row).

\begin{figure}
\centering
\begin{minipage}{.4\textwidth}
\vspace*{-0.1\textwidth}
  \centering
  \includegraphics[width=\textwidth]{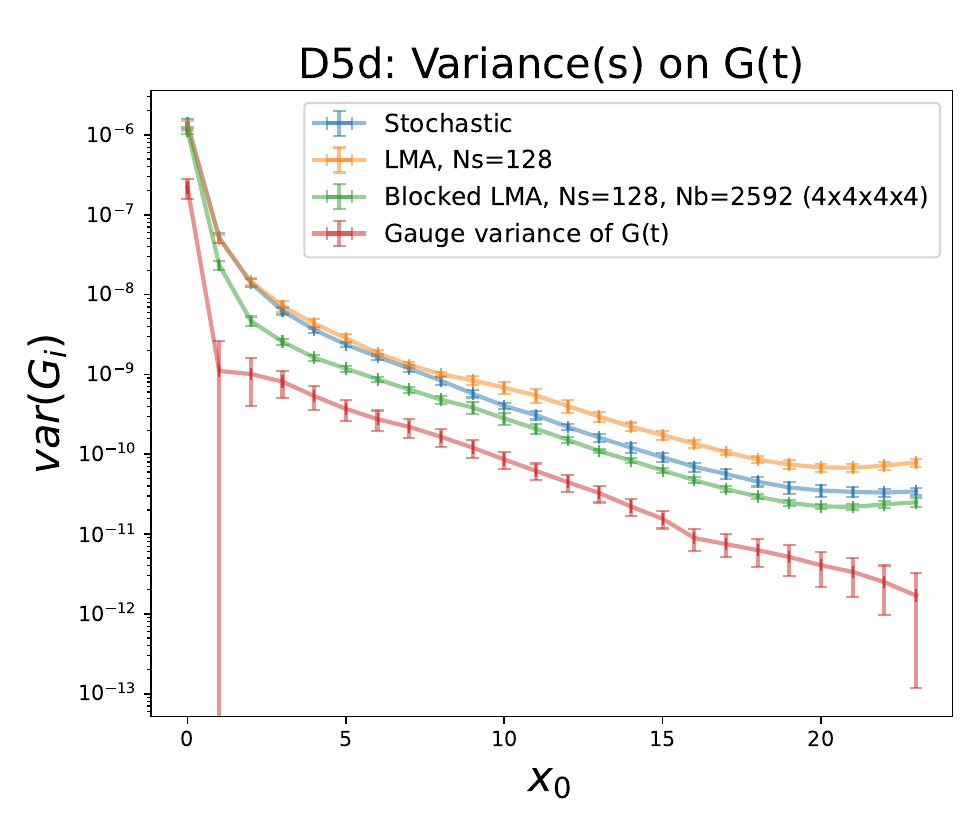}
\end{minipage}%
\begin{minipage}{.4\textwidth}
\vspace*{-0.1\textwidth}
  \centering
  \includegraphics[width=\textwidth]{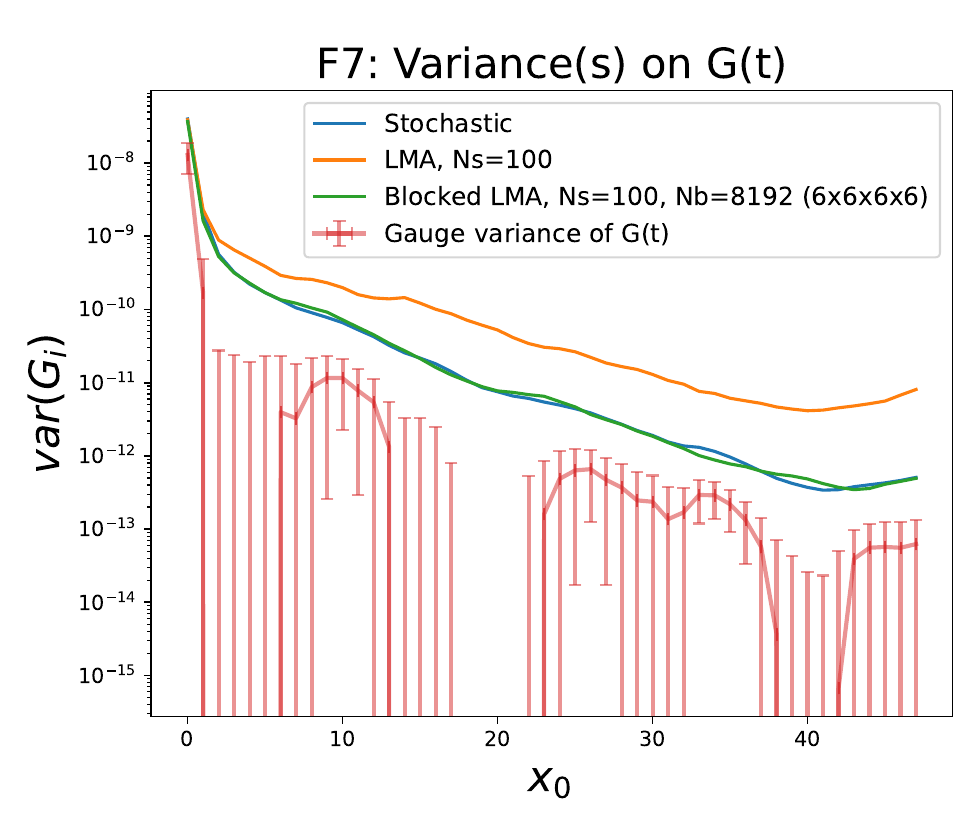}
\end{minipage}

\caption{Comparison of variances of the one-end trick (blue), deflated (yellow)
and deflation with domain-decomposed subspace estimators (green). An estimator
for the variance of the exact translation average Eq.~\eqref{eq:vol} is shown
in red. D5d (left) and F7 (right) with $\Ns=128,100$ low modes and
$\Nsrc=32,96$ stochastic spin-diagonal sources, respectively.}
\vspace*{-0.01\textwidth}
\label{fig:final}
\end{figure}

\section{Conclusions}
In this work we presented a preliminary investigation into variance reduction
using deflation with domain-decomposed subspaces for the isovector vector
correlator with $\Nf=2$ non-perturbatively $\mathrm O(a)$-improved Wilson
fermions.
Such a domain decomposition exploits the local coherence of the low quark modes
of the Dirac operator that almost eliminates the volume-squared problem of
generating such subspaces.
Variance reduction methods are required for the vector channel due to the
large contribution to the variance from the noise fields, as can be seen
in Fig.~\ref{fig:final} where we computed an estimator for the variance of the
translation-averaged observable Eq.~\eqref{eq:vol}.

For the few subspace configurations that we investigated, we observed that in
many cases deflating the stochastic estimator may actually increase the
variance.
In Fig.~\ref{fig:final} (left) for the ensemble with larger quark-mass, we see
a moderate reduction in the variance with the best parameter choice (green)
over the one-end trick (blue).
For the smaller quark mass ensemble (right), the variance is at best identical
to the case of the one-end trick.
We note that in traditional low-mode averaging (yellow), the variance is actually
increased with respect to the one-end trick estimator (blue) with these
parameter choices and for this observable.
Therefore, in order to be safe, a careful analysis of the variances can be is
important to gauge the efficiency of deflation.
Clearly, there are many possible choices to further improve the estimation of
the remainder term (particularly the cross term) which was computed in a simple
way in this work, which is ongoing, as well as work towards a more efficient
implementation.
A more thorough theoretical analysis of low-mode deflation on the variances
would be beneficial to better motivate the definition of the subspace.

\acknowledgments

We are grateful to R. Brower, J. Coles, M. L\"uscher, and the RC* collaboration members for inspiring discussions the interesting discussions, and the feedback on the initial version of this work. We acknowledge the access to Piz Daint at the Swiss National Supercomputing Centre, Switzerland under the ETHZ’s share with the project IDs go24, eth8 and c21. The supported by the PASC project "Efficient QCD+QED Simulations with openQ*D software" is gratefully acknowledged. 

\printbibliography



\end{document}